\documentclass[a4paper,fleqn,usenatbib]{mnras}
\usepackage{newtxtext,newtxmath}
\usepackage[T1]{fontenc}
\usepackage{ae,aecompl}

\usepackage{graphicx}	% Including figure files
\usepackage{amsmath}	% Advanced maths commands
\usepackage{amssymb}	% Extra maths symbols
\usepackage{wasysym}

\title[Simulated Early Star Formation in UFDs]{Exploring Simulated Early Star Formation in the Context of the Ultrafaint Dwarf Galaxies}

\author[L. Corlies et al.]{
Lauren Corlies,$^{1,2}$\thanks{E-mail: lcorlie1@jhu.edu}
Kathryn V. Johnston,$^{2}$ and 
John H. Wise$^{3}$
\\
$^{1}$Department of Physics \& Astronomy, Johns Hopkins University, Baltimore, MD, 21201,USA\\
$^{2}$Department of Astronomy, Columbia University, New York, NY, 10027, USA\\
$^{3}$Center for Relativistic Astrophysics, Georgia Institute of Technology, Atlanta, GA, 30332, USA\\
}

\date{Accepted XXX. Received YYY; in original form ZZZ}

\pubyear{2017}

\begin{document}
\label{firstpage}
\pagerange{\pageref{firstpage}--\pageref{lastpage}}
\maketitle

\begin{abstract}
Ultrafaint dwarf galaxies (UFDs) are typically assumed to have simple, stellar populations with star formation ending at reionization.  Yet as the observations of these galaxies continue to improve, their star formation histories (SFHs) are revealed to be more complicated than previously thought.  In this paper, we study how star formation, chemical enrichment, and mixing proceed in small, dark matter halos at early times using a high-resolution, cosmological, hydrodynamical simulation.  The goals are to inform the future use of analytic models and to explore observable properties of the simulated halos in the context of UFD data.  Specifically, we look at analytic approaches that might inform metal enrichment within and beyond small galaxies in the early Universe.  We find that simple assumptions for modeling the extent of supernova-driven winds agree with the simulation on average whereas inhomogeneous mixing and gas flows have a large effect on the spread in simulated stellar metallicities.  In the context of the UFDs, this work demonstrates that simulations can form halos with a complex SFH and a large spread in the metallicity distribution function within a few hundred Myr in the early Universe.  In particular, bursty and continuous star formation are seen in the simulation and both scenarios have been argued from the data.  Spreads in the simulated metallicities, however remain too narrow and too metal-rich when compared to the UFDs.  Future work is needed to help reduce these discrepancies and advance our interpretation of the data.  

\end{abstract}


\section{Introduction} \label{sect.intro}
Since their discovery, dwarf galaxies in the Local Group have been thought to probe the early Universe and to contribute some of the oldest stellar populations to the larger Milky Way galaxy \citep{dekel_1986,diemand_2005,font_2006,frebel_2012}.  In particular, the smallest of these, the ultrafaint dwarf galaxies (UFDs), were proposed to host simple, stellar populations with star formation histories (SFHs) that were truncated by reionization \citep[][for a review]{bovill_2009, salvadori_2009, tolstoy_2009}.  Their overabundance of $\alpha$-elements relative to their metallicity is one signature of this early and short star formation \citep{kirby_2011,vargas_2013}.

However, the SFHs of these small objects may be more complex than first thought.  Deep colour-magnitude diagrams (CMDs) and broad metallicity distribution functions (MDFs) of six UFDs suggest that all of their stars have ancient ages (forming more than 10.5 Gyr ago) as expected, but that in many cases, the SFHs can be best fit with a two-burst model.  Yet, a single burst model can not be ruled out in any case and in some is in fact the best fit \citep{brown_2014}.  When including observations of the $\alpha$-abundance in such SFH fits, the stars with higher metallicities in the UFDs sometimes also have lower [$\alpha$/Fe], a signature of enrichment of Type Ia supernovae (SNe) and chemical self-enrichment in each galaxy \citep{webster_2015}.  The time needed for Type Ia SNe to explode and enrich surrounding gas in turn implies a more extended and continuous SFH for the UFDs.  One exception to this is Segue 1, whose MDF has distinct peaks that can be described with discrete bursts of star formation \citep{webster_2016}.  Thus, a more complete picture of the early star formation in these small galaxies can be better reconstructed by combining the CMDs, MDFs, and $\alpha$-abundances.

Generally, interpreting such data to reconstruct detailed SFHs relies on an underlying chemical evolution model.  Using simple yet powerful parameterizations, these models can trace the build up of iron and other elements as stars form within a galaxy of a given mass, creating tracks in abundance space that recreate what is seen in studies of Milky Way halo stars \citep{mcwilliam_1997}.  Whether purely analytic \citep{robertson_2005, andrews_2017}, coupled to an N-body simulation \citep{font_2006, tumlinson_2010, romano_2013, crosby_2016}, or incorporating global, measured SFRs \citep{lanfranchi_2008,avilavergara_2016}, these models allow for quick variation of their parameters and for an assessment of the importance of a number of galaxy properties (e.g. inflows, outflows, SFH) in driving the chemical evolution of a given halo.  However, an underlying assumption of all of these models is the complete mixing of metals within a halo once returned from stars to the modeled ISM, whether instantaneously or within a parametric timescale.  Furthermore, halos are typically considered to be chemically isolated as they evolve.  Both of these assumptions limit any model's ability to reproduce the \emph{scatter} seen in the measured stellar abundances. One exception to this is the set of models where SNe are considered to have individual mixing volumes which can then overlap as they expand \citep{oey_2003, karlsson_2005,leaman_2012,gomez_2012}.  %ADD DILUTION MASS HERE ??

Thus, while these chemical evolution models are flexible and powerful, there are some inherent limitations.  The natural solution is to turn to hydrodynamical simulations.  Idealized simulations of an UFD-sized galaxy have been constructed to study the influences of radiative cooling, clumpy media, and off-centered explosions - effects typically not included in analytic models \citep{blandhawthorn_2015}.  Such simulations can also be used to test complex explosion scenarios to reproduce specific SFHs and MDFs \citep{webster_2015, webster_2016}.  For cosmological simulations, it unfortunately remains computationally prohibitive to simulate the formation and evolution of a UFD around a Milky Way-like galaxy to $z=0$ because of the necessarily high resolution.  New simulations have begun to identify such UFDs and simulate their evolution to $z=0$ while even tracking individual elemental abundances in star particles but still at large distances from the host halo and still leaving the massive host less resolved \citep{jeon_2017}.  Typically though, simulations have focused on isolated dwarf galaxies, finding SN feedback and UV background radiation \citep{sawala_2010} as well as the time of reionization \citep{simpson_2013} are critical for reproducing galaxies with characteristics similar to the Local Group dwarf spheroidals.  On the other hand, a larger galaxy population can be simulated at high resolution if only high redshifts are considered.  Such work emphasizes the importance of SN feedback in providing turbulent-driven mixing \citep{ritter_2015}, external enrichment from nearby halos \citep{smith_2015}, and providing a metallicity floor for subsequent star formation\citep{wise_sims} - all typically considered in the context of Population III (Pop III) star formation.  The beginning of Population II (Pop II) star formation and the impact of these stars on the global metallicity evolution of early galaxies is an area ready for exploration with this new class of high-resolution simulations.

Simultaneously, the population of observed dwarf galaxies is again expanding as the Dark Energy Survey begins its comprehensive search of the southern sky.  Seventeen new dwarf galaxy candidates have been found in the survey's first two years \citep[see][and the references within for a summary of the observations]{des_dwarfs_2015}.  Since observations of small galaxies in the early Universe will remain difficult even with JWST \citep{okrochkov_2010} and because the Local Group is thought to be a representative cosmic volume at early times \citep{boylan-kolchin_2016}, Local Group UFDs will remain as the basis for near-field cosmology to answer fundamental questions about galaxy formation.  Their numbers and SFHs can already provide constraints on when and how reionization proceeded \citep{bullock_2000, benson_2002, busha_2010}; suggest that they were primary sources of chemical enrichment for the intergalactic medium (IGM) at high redshift \citep{scannapieco_2002, salvadori_2014}; and place constraints on the potential nature of warm dark matter \citep{chau_2017}.  Furthering our understanding of these objects will build a more complete picture of how galaxies formed globally.

In summary, observations suggest that the UFDs have short but complex SFHs and that relaxing some basic assumptions about homogeneous mixing and isolated environments may play a role in creating the spreads seen in abundance patterns.  This paper examines a high resolution, cosmological simulation of the early Universe to see how star formation, enrichment, and mixing proceed in small dark matter halos in order to inform the use of analytic models and to consider what the observational consequences might be.  In Section \ref{sect-sims}, the simulation being analyzed is described in detail.  In Section \ref{sect-dwarfs}, a summary of the characteristics of the simulated dwarf galaxies that are the basis of this paper is presented.  In Section \ref{sect-models}, we assess the validity of two common assumptions of analytic models, homogeneous mixing and isolated environments, by comparing models to the hydrodynamical simulation where the assumptions are relaxed.  In Section \ref{sect-obs}, we present the SFHs and MDFs of the simulated halos for comparison with the observations.  Finally, we discuss the implications of these results in the context of current simulations in Section \ref{sect-discuss} and summarize our conclusions in Section \ref{sect-conc}.

\section{Simulation Basics} \label{sect-sims}
We analyze the cosmological, radiation, hydrodynamical simulation of \citet{wise_rad} (referred to as ``RP'' in the paper), performed with {\sc Enzo}, an Eulerian, adaptive mesh refinement, hydrodynamical code \citep{enzo}.  The simulation has a box size of 1 Mpc and a resolution of $256^3$ leading to a dark matter particle with $m_{\mathrm{DM}} = 1840 M_{\astrosun}$.  This mass resolution coupled with a maximal spatial resolution of 1 co-moving pc enables detailed studies of the formation of the first generation of dwarf galaxies.  The simulation was stopped at $z=7$ to prevent any large-scale modes with $r \approx L_{\mathrm{box}}/2$ from entering the non-linear growth regime. 

The simulation accounts for both the formation and metal yields of Pop II and III stars separately.  Pop II stars are formed if [Z/H] $> -4$ and Pop III stars are formed otherwise.  Star particles are formed if the gas has an overdensity of $5 \times 10^5 (\approx 10^3 \mathrm{cm}^{-3}$ at $z=10$) and has a converging flow.  Pop III stars also have an additional formation requirement that the gas has a molecular hydrogen fraction such that $f_{\mathrm{H}_2} > 5 \times 10^{-4}$.  Instead of forming star particles with a fixed stellar mass, Pop III stellar masses are sampled from an IMF with a functional form of: 

\begin{equation}
f(\log M )dM = M^{-1.3} \exp \left [ - \left ( \frac{M_{\mathrm{char}}}{M} \right )^{1.6} \right ] dM
\end{equation}
where $M_{\mathrm{char}} = 100 \ M_{\astrosun}$.  Above $M_{\mathrm{char}}$, it behaves as a Salpeter IMF but is exponentially cutoff below \citep{chabrier_2003,clark_2009}. In this way, each Pop III particle represents an individual star.

Pop II stars follow the same formation criteria with the minimum H$_2$ fraction requirement relaxed. Once a cell meets the criteria, a molecular cloud is defined by setting the densest point as the cloud center and searching outwards until a sphere is found with a dynamical time  $t_{\mathrm{dyn}} = 3$ Myr (corresponding to an average density $\bar{\rho} \simeq 1000 \mu \  \mathrm{cm}^{-3}$) and a radius $R_{\mathrm{MC}}$, where $\mu$ is the mean molecular weight.  With the sphere identified, a mass fraction of the cold gas ($f_{\mathrm{cold}}$, $T < 10^3$ K) is converted into a star particle with mass $m_* = 0.07 f_{\mathrm{cold}} (4\pi/3) \bar{\rho}_{\mathrm{MC}}  R^3_{\mathrm{MC}}$.

The minimum mass of a Pop II star particle is $1000 \  M_{\astrosun}$, representing an entire stellar population. Most star particles are formed above this limit and immediately emit ionizing radiation.  For cells that meet the star formation criteria but are below this limit, a sink particle is placed and allowed to accrete cold gas for up to 3 million years. Once this mass is reached, the star particle stops accreting and is activated. If it does not reach this mass limit within the timeframe, the accretion phase is terminated and begins to emit ionizing radiation. This results in star particles with a range of final masses.

The SN feedback is included in the simulation as thermal feedback.  The blast wave is modeled by injecting the explosion energy and ejecta mass into a sphere of 10 pc, smoothed at its surface to improve numerical stability.  Because the simulation resolves the blast wave relatively well with several cells at its initialization, the thermal energy is converted into kinetic energy and agrees with the Sedov-Taylor solution \citep[e.g. ][]{greif_2007}.  Pop III can form three types of supernovae: normal Type II SNe ($11 \leq M_*/M_{\astrosun} \leq 20$), hypernovae ($20 \leq M_*/M_{\astrosun} \leq 40$) \citep{woosley_1995}, and pair-instability supernovae (PISNe, $140 \leq M_*/M_{\astrosun} \leq 260 $) \citep{heger_2002}.

For Pop III star particles within the Type II SNe mass range, an explosion energy of $10^{51}$ erg is used and a linear fit to the metal ejecta mass is calculated as in \citet{nomoto_2006}:

\begin{equation}
M_Z/M_{\astrosun} = 0.1077 + 0.3383 \times (M_* / M_{\astrosun} - 11)
\end{equation}

Pop III hypernovae are an extension of the \citet{nomoto_2006} model, linearly interpolating results to $M_*$.  For Pop III PISNe, the following function is fit to the models of \citet{heger_2002}:

\begin{equation}
E_{\mathrm{PISN}} = 10^{51} \times [5.0 + 1.304(M_{\mathrm{He}}/M_{\astrosun} -64)] \ \mathrm{erg}
\end{equation}
where $M_{\mathrm{He}} = (13/24) \times (M_* - 20)M_{\astrosun}$ is the helium core (and equivalently the metal ejecta mass) and $M_*$ is the stellar mass.  If the star particle mass is outside of these ranges, than an inert, collisionless black hole particle is created.

Finally, Pop II star particles generate thermal SNe feedback with energy $6.8 \times 10^{48}$ erg s$^{-1}$ $M_{\astrosun}^{-1}$ after living for 4 Myr.  The ejected gas has solar metallicity, $Z = 0.02$, resulting in a total metal yield of $y=0.005$.  This yield results from assuming that, given a Salpeter initial mass function, that $25\%$ of the mass is returned to the ISM whereas the remaining metals are locked up in stellar remnants and low mass stars that never become SNe.

In addition to this thermal feedback, a unique feature of the simulation is its inclusion of the energy coupling of the radiation field generated by these stars to the gas surrounding them.  The radiation field is followed with adaptive ray tracing \citep{wise_2011} that is based on the HEALPix framework \citep{gorski_2005} and is coupled self-consistently to the hydrodynamics.  The radiation is modeled with an energy, $E_{\mathrm{ph}}$, equaling the luminosity-weighted photon energy of the spectrum. Pop III stars have a mass-dependent luminosity taken from \citet{schaerer_2002}, with $E_{\mathrm{ph}} = 29.6$ eV, appropriate for the near-constant $10^5$ K surface temperatures of such stars.  Each Pop III star particle then dies with an endpoint according to its mass as described above. For Pop II stars, 6000 photons per baryon with $E_{\mathrm{ph}} = 21.6$ eV are emitted, appropriate for stars with [Z/H] = -1.3.

For more details, we direct the reader to Section 2 of \citet{wise_rad}.  The analysis and plots that follow were done with {\sc yt} \citep{yt}.

\section{Simulated Dwarf Galaxy Population} \label{sect-dwarfs}
\begin{figure}
\centering
\includegraphics[width=0.4\textwidth]{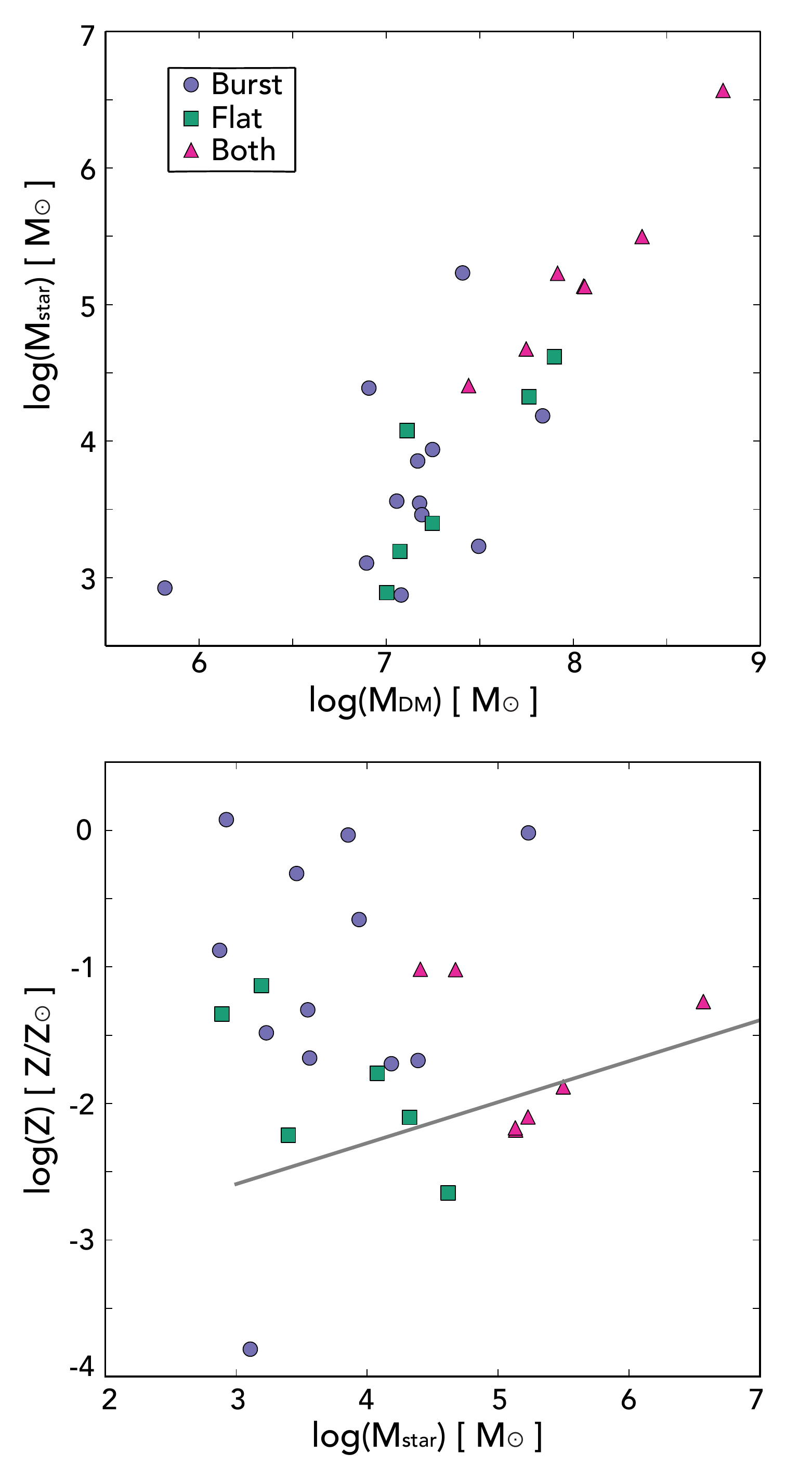}
\caption{Upper panel: Stellar mass as a function of total dark matter halo mass showing the range of halos in the simulation. Lower panel: stellar mass versus average metallicity ([Z/H]) for the simulated star-forming halos. The labels correspond to SFH types. (see Section 4.1) The line corresponds to the observed relation for Local Group dwarf galaxies \citep{kirby_2013}.  While the simulated galaxies have roughly the same masses as the observed dwarfs, they are generally too metal rich and there is a much higher scatter in their metallicities.  \label{massZ.fig}}
\end{figure}

Even in the early universe of the simulation, a number of halos form stars and represent a range of dark matter masses, stellar masses, and metallicities.  The full population will be discussed throughout.  Of the almost 2000 halos at $z=7$ within the simulation with $M_{\mathrm{halo}} > 5 \times 10^5 M_{\astrosun}$, 24 of them have formed stars by the final snapshot.  Their distribution of halo masses, stellar masses, and average metallicities can be found in Figure \ref{massZ.fig}.  They range from roughly $10^5$ M$_{\astrosun}$ - $10^8$ M$_{\astrosun}$ in halo mass, $10^3$ M$_{\astrosun}$ - $10^6$ M$_{\astrosun}$ in stellar mass, and $-2.5 <$ [Z/H] $ < 0.1$ in average metallicity. Star formation proceeds in these halos for a period of roughly 500 Myr where it can no longer be followed as the simulation is terminated because of computational costs. 

To place this dwarf population in the context of the observed dwarf galaxies.  Figure \ref{massZ.fig} shows the mass-metallicity relation for the simulated halos as well as the observed relation of \citet{kirby_2013}.  Seven already have $M_* > 10^4 M_{\astrosun}$, the standard cutoff for a UFD classification at lower redshift \citep{martin_2008}.  However, when the entire mass range of the observed Local Group dwarfs is considered, the simulated dwarfs span roughly the same mass range.  Yet the halos within the simulation typically have higher average metallicities than their low-$z$ Local Group counterparts.  We consider the implications of this in Section 5.2.

\section{Exploring Analytic Model Assumptions} \label{sect-models}
Analytic models remain the fastest and most flexible way to evaluate the importance of different baryon properties in the smallest halos at $z=0$.  This section focuses on the role of inhomogeneous mixing and chemical isolation in the simulation and what deviations from analytic models these processes can create. 

\subsection{Closed Box Model and Inhomogeneous Mixing}
The simplest model for tracing the metallicity of the gas and stars associated with a dark matter halo is the closed box model.  It assumes that there is no flow of gas into or out of the halo and that the gas self-enriches with each star formation event.  Furthermore, the amount of metals that is returned to the gas is determined by the assumed yield, the mass of metals formed by a given stellar mass, and the metals are assumed to mix completely and instantaneously.  Within this simple framework, the metallicity is given analytically as:

\begin{eqnarray}
Z(t) &=& -p \ln \left ( \frac{M_{\mathrm{gas}}(t) }{M_{\mathrm{gas}}(0)} \right ) \\ 
       &= &-p \ln \left( \frac{M_{\mathrm{gas}}(t)}{ M_{\mathrm{gas}}(t) + M_{\mathrm{star}}(t) } \right ) \nonumber
\end{eqnarray}
where $p$ is the yield, $M_{\mathrm{gas}}(0)$ is the initial gas mass, and $M_{\mathrm{gas}}(t)$ and $M_{\mathrm{star}}(t)$ are the current gas and stellar mass respectively.  Deviations from this model in the simulation can then be used to discuss the role of gas flows and mixing within the halos. 

\begin{figure}
\centering
\includegraphics[width=0.5\textwidth]{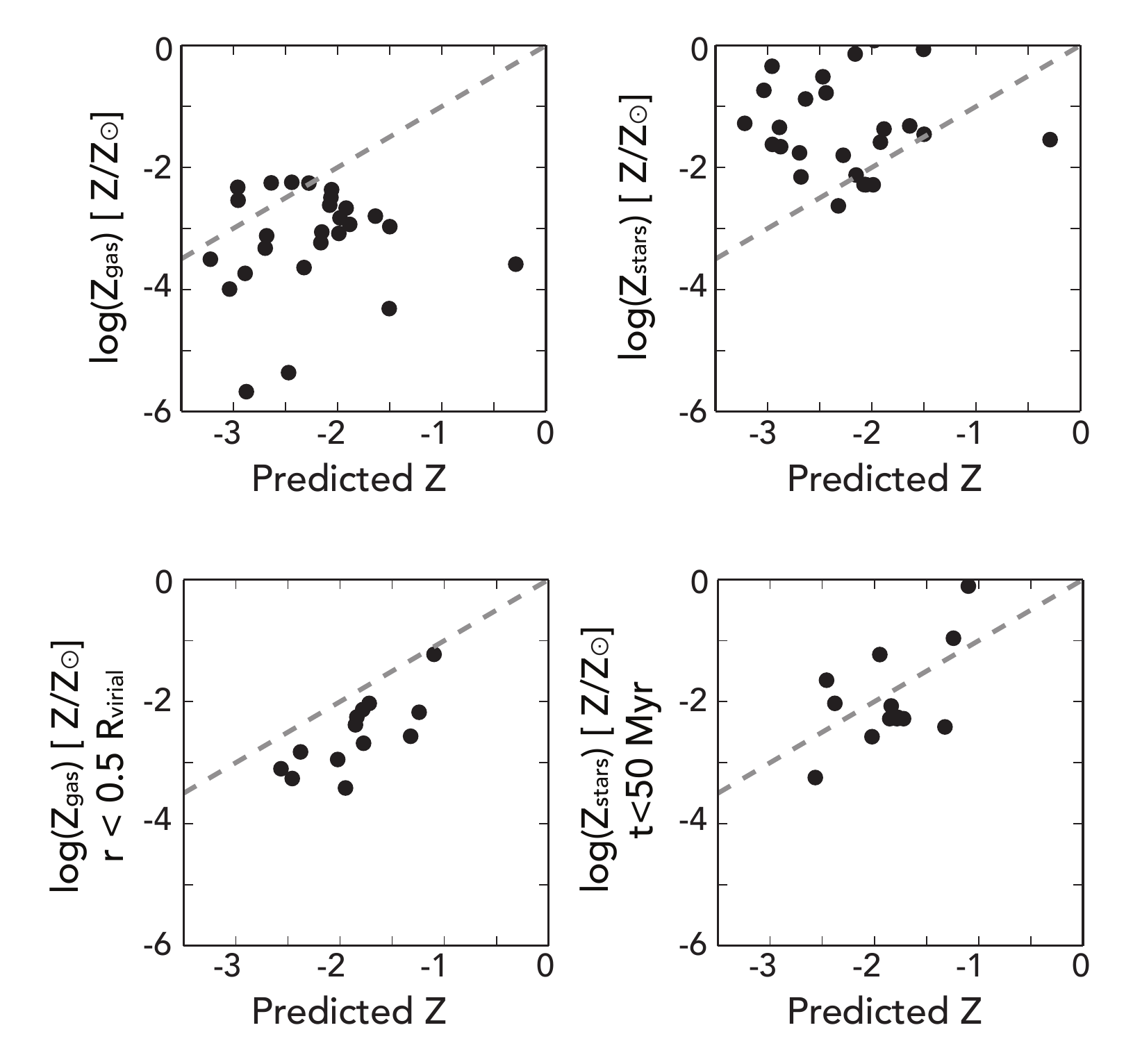}
\caption{Top panels: closed box predicted metallicities versus the volume-weighted average gas metallicity (left) and particle mass-weighted stellar metallicities (right) within the virial radius of each halo at $z=7$.  The dashed line shows where the two quantities would be equal. The large spread in metallicities as well as the discrepancy between the gas and stellar metallicities indicate that the metals are not fully mixed within the halo.  Lower panels: closed box predicted metallicities compared to the average gas metallicity within half the virial radius (left) and to the average metallicities of stars formed within the last 50 Myr of z=7 (right) for each halo.  There are fewer points because only a fraction of the full sample has formed stars within this time frame. Restricting the volume considered brings the gas metallicities into better agreement with the model.  Similarly, limiting the stars to those that formed more directly from the gas at the considered redshift brings better agreement with the model.  \label{closedbox.fig}}
\end{figure}

Figure \ref{closedbox.fig} shows how the metallicities predicted by this model compare with those seen in the simulations at $z=7$. For each star-forming halo, the final gas and stellar masses within the virial radius are used to compute a closed box-predicted metallicity using Equation 4.  The simulated gas metallicities are volume-weighted averages while the stellar metallicities are particle-mass-weighted averages, within the virial radius and at the same redshift.  As seen in the top left panel, the simulated gas metallicities are mostly lower than model predictions where we would expect them to be equal if the model accurately described the simulated galaxies.  In contrast, the stellar metallicities (top right panel) are almost entirely higher than what is predicted by the model.  Yet because this is the full stellar population, we might anticipate that the metallicities be lower than the model predicted values because the averages here include early formed, low-metallicity stars and not just stars formed out of the higher enriched gas at the final redshift.  Introducing a simple offset in the model (for example by changing the yield) may shift its values up or down in Figure \ref{closedbox.fig} but would not be enough to reproduce the large scatter seen in the gas and stellar metallicities of the halos. 

Most analytic models, including the closed box model, consider global halo properties, like the total gas and stellar masses.  Yet in the simulation, stars are formed in the center parts of halos where the gas is the densest and only the most recently formed stars should be connected to the current state of the gas.  To this end, the bottom left panel of Figure \ref{closedbox.fig} shows how the closed box model compares to the simulated metallicities when only gas within $50\%$ of the virial radius is considered.  Similarly, the bottom right panel takes the closed box metallicity and compares it to the average metallicity of stars formed within 50 Myr of the last time step.  Only a fraction of the full sample has formed stars within this time frame, resulting in a smaller number of points for these panels. Physically, this inner gas volume should be directly related to the most recently formed stars of each halo and we do indeed see that the simulated gas and stellar values now fall much closer to each other and to the values predicted by the closed box model.  While a better agreement from recently formed stars to the current state of the gas was expected, it is less clear why restricting the volume considered for the average gas metallicity brings the simulation into much better agreement with the closed box model, when the model is intended to describe the global gas properties of the halo.

\begin{figure*}
\centering
\includegraphics[width=0.75\textwidth]{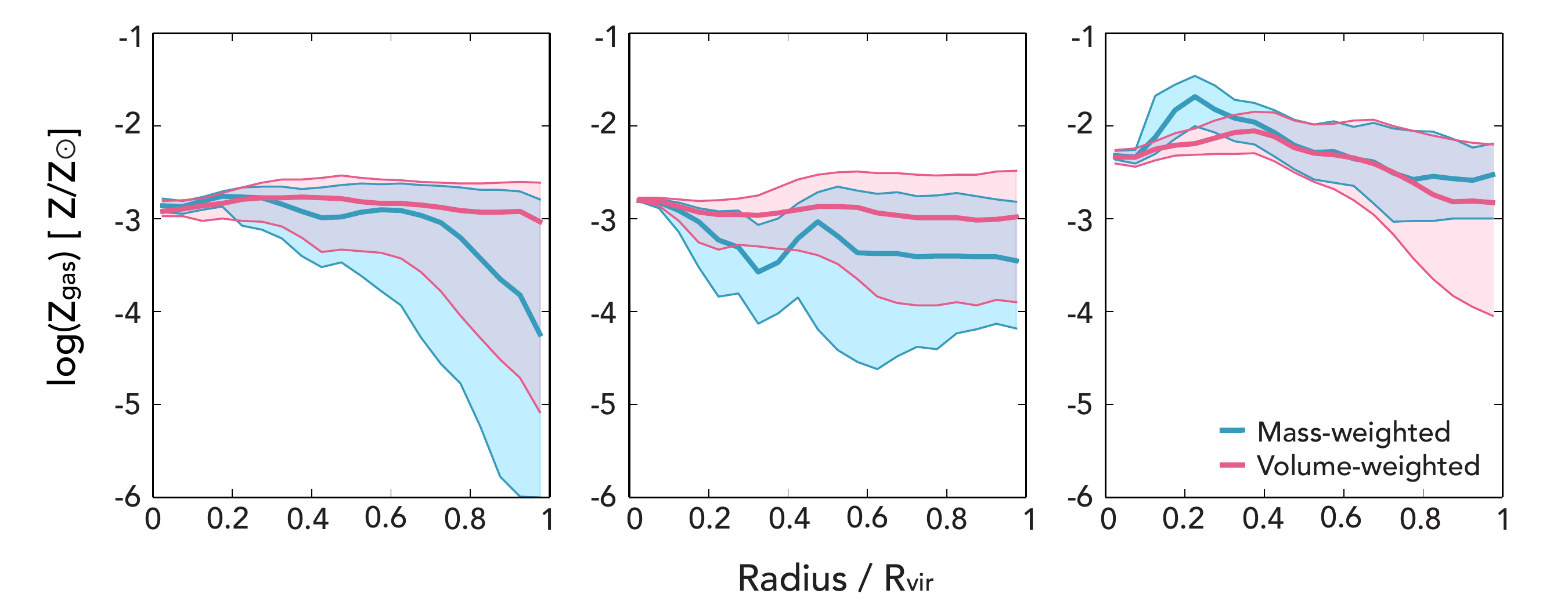}
\caption{Radial profiles of the gas metallicity for three simulated halos at $z=7$.  Lines show the average and shaded regions show the 25th and 75th percentiles for the mass-weighted (blue) and volume-weighted (pink) metallicities.  The central parts of the halos are more well-mixed, and may experience a better balance between inflows and outflows.  The outskirts, on the other hand, show much larger spreads in metallicity and a mismatch between the mass-weighted and volume-weighted values.  \label{radprofs.fig}}
\end{figure*}

To try and answer this question, Figure \ref{radprofs.fig} shows the radial profiles of the volume-weighted (pink) and mass-weighted (blue) gas metallicities for three halos within the simulation.  These halos are representative of the entire halo population.  The lines represent the median value within a given spherical shell while the shaded regions represent the 25th and 75th percentiles.  For all the halos, the profiles within half the virial radius are much narrower and more metal-rich.  The central parts of the halos are more well-mixed, and may experience a better balance between inflows and outflows, explaining their better agreement with the closed box model.  The outskirts, on the other hand, show much larger spreads in metallicity and a mismatch between the mass-weighted and volume-weighted values.  Inflowing gas is denser and more metal poor, skewing the mass-weighted values low.  Outflowing gas is more diffuse, more metal rich, and more volume-filling, skewing the volume-weighted averages high.  Finally, at no radii is the gas fully mixed.

In summary, the closed box model can be surprisingly accurate in representing the average halo gas metallicity, if the volume in question is limited to within $50\%$ of the virial radius, despite all of its simplifying assumptions being violated.  However, as Figure \ref{radprofs.fig} demonstrates, this does not capture the true metallicity properties of the halos, which show variations with radii in both the median metallicity and the amount to which the gas is well mixed.  Any understanding of the observed spread in metallicities will require more accurate modeling of inflows, outflows, and mixing within a given galaxy.

\subsection{Supernova-Driven Winds and Chemical Isolation}
SNe are known to drive winds from galaxies that enrich their surroundings \citep{heckman_2000, aguirre_2001, martin_2013} and other nearby halos \citep{wise_sims, smith_2015}.  This outflow process can be approximated as a pressure-driven spherical shell, expanding with the Hubble flow and into an IGM with constant density ($\bar{\rho}$) of zero pressure \citep{tegmark}. In this simple configuration, the radius of the outflow evolves according to the following equation of motion:

\begin{equation}
	    \ddot{R}_s = \frac{3P_b}{\bar{\rho}R_s} - \frac{3}{R_s}(\dot{R}_s - HR_s)^2 - \Omega_m \frac{H^2R_s}{2}
	\end{equation}
where the overdots represent time derivatives and the $s$ and $b$ subscripts indicate shell and bubble quantities respectively. The pressure, $P_b$, that drives the bubble is provided by the SNe, which have a net input of energy into the system with a rate equal to
\begin{equation}
	\dot{E}_b = L(t) - 4\pi R^2_s \dot{R}_sP_b
\end{equation}
Here, $L(t)$ is the energy injection from SNe (originally referred to as luminosity in \citet{tegmark} and continued here) and the remaining term is the typical work done by the shell as it expands.  Lastly, adiabatic expansion is assumed such that $P_b = E_b/2 \pi R^3_s$.

The luminosity function above is set by the given SFH.  If the gas mass is assumed to be a constant fraction, $f_{\mathrm{bary}}$, of the dark matter mass, then in any time interval, $\Delta t$, the mass of stars formed is:

\begin{equation}
M_{\mathrm{star}} = \frac{ f_{\mathrm{bary}} \ M_{\mathrm{DM}} \ \Delta t }{ \tau_{\mathrm{sf}} }
\end{equation}
where $f_{\mathrm{bary}} = 0.05$ and $\tau_{\mathrm{sf}}$, the star formation timescale, is 10 Gyr.  These parameters were found to well reproduce the total stellar masses as well as the stellar mass density profiles of a Milky Way-like galaxy and its low mass satellites by \citet{tumlinson_2010}, which is important for our comparison with the local UFDs. This low gas fraction is also supported by scaling relations seen in high-$z$ cosmological simulations \citep{chen_2014}.   

Finally, following \citet{tumlinson_2010}, it is assumed that 1 SN occurs for every 100 $M_{\astrosun}$ sun of star formation for a \citet{kroupa_2001} IMF and that each SN produces $10^{51}$ ergs of energy.  Thus, the SNe luminosity can be defined as:

\begin{equation}
L(t) = (0.01 \  \mathrm{SNe}/M_{\astrosun})(10^{51} \ \mathrm{erg/SNe}) \left( \frac{ f_{\mathrm{bary}} M_{\mathrm{DM}}}{  \tau_{\mathrm{sf}} } \right)
\end{equation}

Levels of complexity (such as different forms of cooling, a stochastic star formation rate, dissipative heating) can be added as desired but there are almost always four underlying assumptions to the model: (1) spherical shells; (2) isolation of the given halo; (3) expansion into a constant density IGM; and (4) conservation of energy as the shell propagates through the IGM.  Understanding that most of these assumptions are violated in a cosmological, hierarchical setting, we measure the extent of the metal-enriched material surrounding each of the star-forming halos in the simulation.  Using {\sc yt}'s clump-finding algorithm, a metal-rich bubble can be identified by setting a metallicity boundary and a search volume without assuming anything about the shape of the bubble.  For our search, we focus on Pop II-driven winds and set a boundary of [Z/H ]$ = -6$ in Pop II-generated metallicity, such that anything with a lower metallicity is considered un-enriched.  The search volume is visually confirmed to be big enough to encompass the bubble associated with each halo.  Once the metallicity surface has been identified, the center of mass of the asymmetric contained volume is defined and the average distance of each cell within the volume to the center is computed and designated as the average radius of the clump.

\begin{figure*}
\centering
\includegraphics[width=0.75\textwidth]{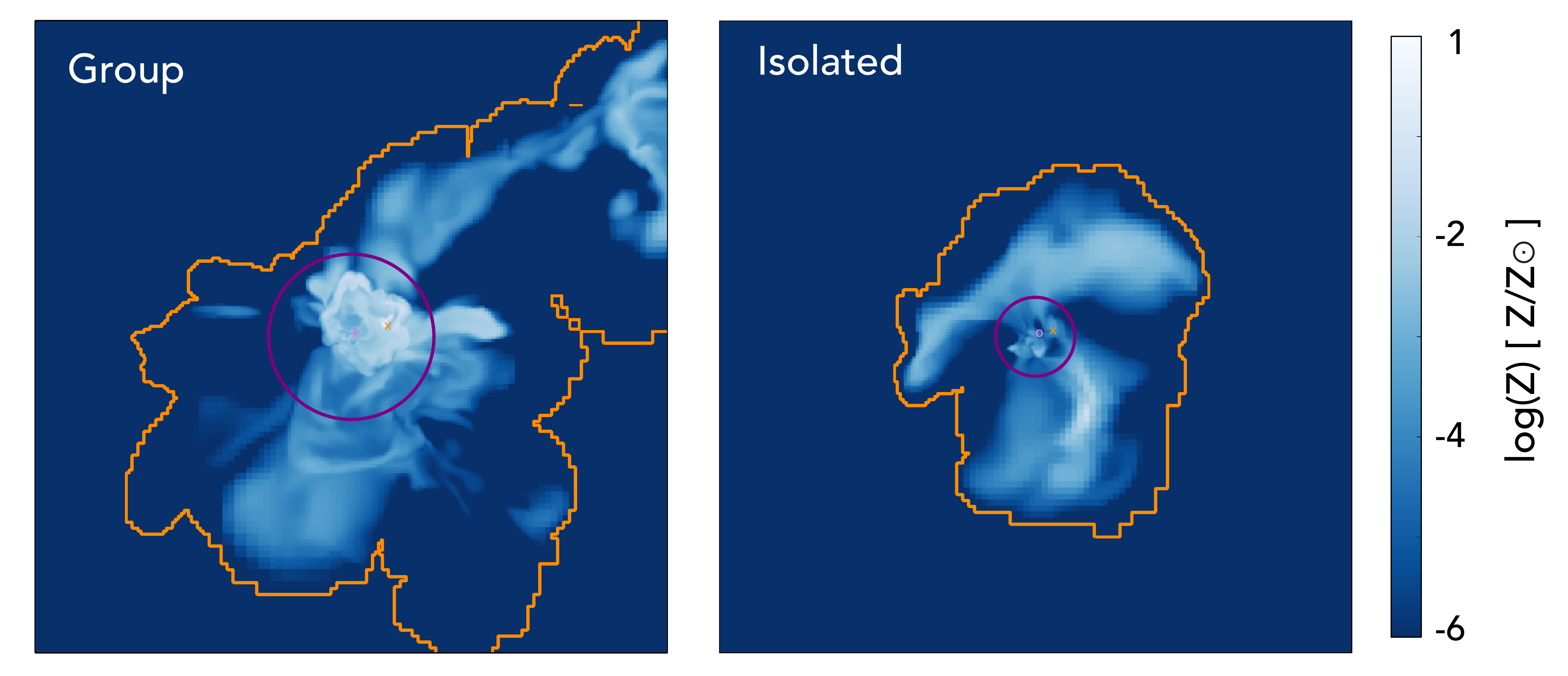}
\caption{Slices through the center of each halo in metallicity with a width of 25 kpc for two different halos at $z=7$ with the bubble contour identified by {\sc yt} shown in orange.  Purple circles represent the virial radii of the halos contained within the simulation slice. The left panel is centered on the most massive halo in the simulation, which is found in a dense, crowded environment. The enrichment bubble surrounding this group of halos is asymmetric and less massive than the semi-analytic model predicts. The right panel shows a more symmetrical bubble for another massive halo in the simulation that is isolated from surrounding halos.   \label{bubbles.fig}}
\end{figure*}

Figure \ref{bubbles.fig} shows examples of this bubble finding for two different halos - the most massive and its surrounding environment on the left and an isolated halo on the right.  A majority of the star-forming halos in the simulation are in a state similar to the most massive halo; that is, most exist within bubble regions that are asymmetrical and encompass multiple halos.  In such a region, the most massive halo can drive a strong wind that is in close enough proximity to cross-enrich the halos surrounding it, violating the basic assumption of analytic models - chemical isolation.  Yet there is a fraction of halos that do appear to remain chemically isolated, driving their own metallicity bubble as shown in the right panel of Figure \ref{bubbles.fig}.  This halo has a more symmetric outflow and can perhaps be expected to be described more accurately by the analytic model of Equations 5-8. 

To facilitate a comparison of the SN-driven wind model to the simulation, we calculate a model-predicted radius for the SN-wind of each halo using the final dark matter halo mass at $z=7$.  The history of the dark matter mass evolution can then be calculated using the formalism of \citet{wechsler_2002}:

\begin{equation}
M_{\mathrm{DM}}(z) = M_0 \exp \left[ \frac{-S}{1 + z_c} \left( \frac{ 1+z }{ 1+z_0 } - 1\right)  \right]
\end{equation}
where $S = 2.0$, $z_0$ is the observed redshift, $M_0$ is the mass at the observed redshift, and $z_c$ is the characteristic formation time.  The starting time for the wind integration was chosen to be the time that the first star particle of the halo was created.  Thus, for each halo, the dark matter halo mass, stellar mass, and wind radii are computed using the model described above, anchored to the final dark matter mass at $z=7$.  A full description of the model can be found in \citet{corlies_2013}. 

\begin{figure}
\centering
\includegraphics[width=0.47\textwidth]{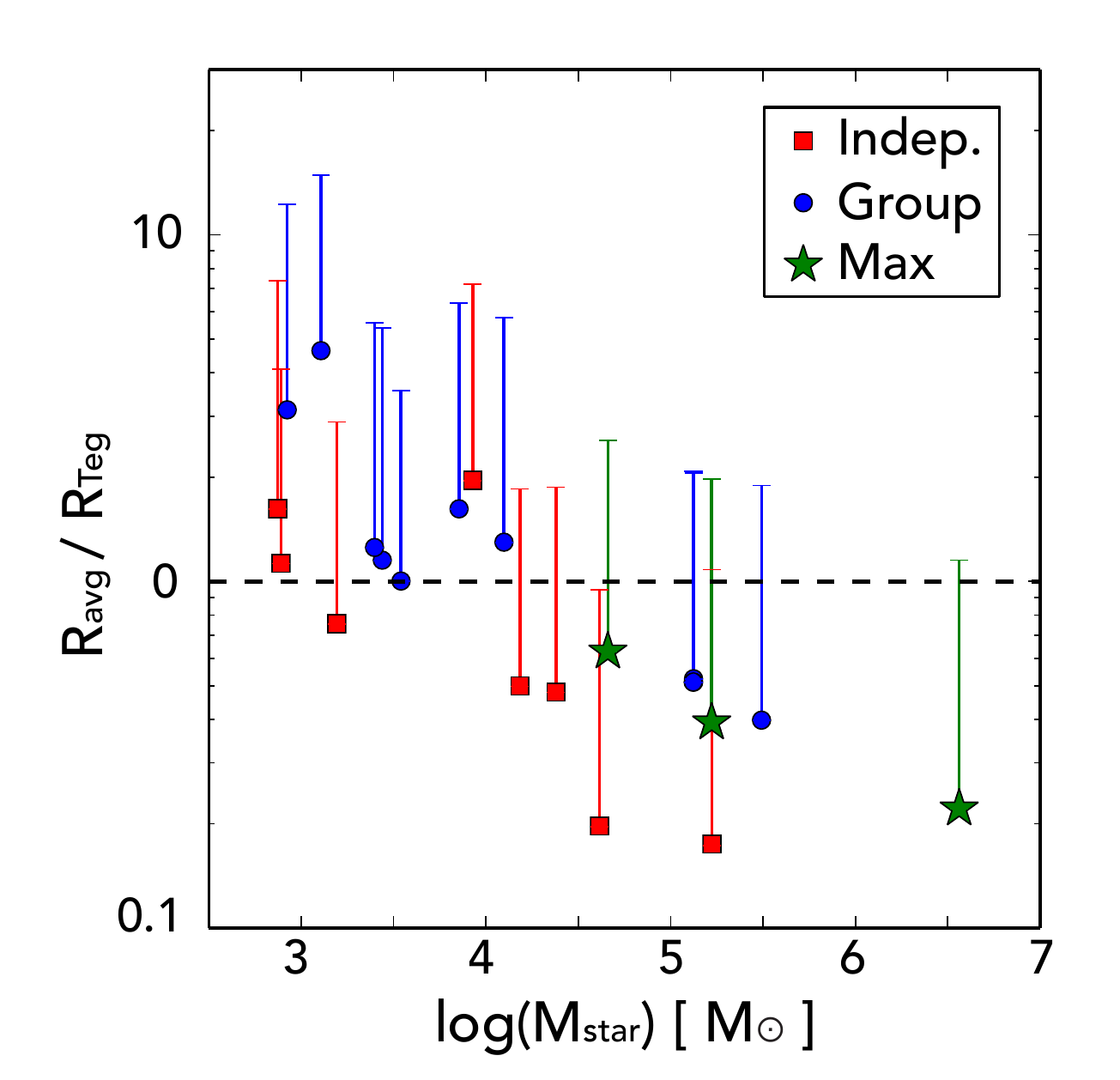}
\caption{Bubble radii predicted by the SAM described in Section 3.2 are compared to those measured by the simulation and plotted as a function of halo stellar mass. Because of their asymmetry, the average bubble radii are plotted as points and the error bar corresponds to the maximum bubble radii associated with each star-forming halo in the simulation.  Isolated halos are plotted as red squares; those sharing enrichment bubbles are plotted as blue circles; and the most massive group member in a bubble is plotted as a green star.  Above a stellar mass of $10^4$ M$_{\astrosun}$, the model radius is within a factor of two of the simulated radius. At lower masses, there is a larger scatter in the ratio of the radii because the bubbles surrounding group members are dominated by their most massive member. Those halos that are isolated show similar agreement between the model and the simulation. \label{tegmark.fig}}
\end{figure}

Figure \ref{tegmark.fig} shows how the average, simulated, bubble radius compares to the model-predicted radius as a function of stellar mass.  Because the bubbles identified by {\sc yt} are asymmetrical, the average radius from the center of mass of the bubble is plotted as a point and the maximum radius is shown at the upper error bar.  In general, as the stellar mass decreases, the ratio of the simulated to modeled bubble radius increases.  However, these quantities never differ too greatly and the higher mass halos are particularly well matched by the model.  This indicates that despite its simplifying assumptions, the analytic model mostly captures the extent of the enrichment seen in the simulation.  

Yet, as noted above, the majority of the halos are in groups, violating the isolation assumption of the analytic model.  To investigate whether a halo's isolation changes its agreement with the model, the halos are separated into three categories.  Those residing alone in their own enrichment bubble are called ``isolated'' while those found sharing enrichment bubbles are identified as either group members (groups') or as the most massive group member (max).  

The most massive halo is expected to drive the strongest wind in the model, so for those halos in groups, the most massive should perhaps correspond most closely to the model value.  Figure \ref{tegmark.fig} shows that the radii of the bubbles associated with the most massive halos are generally under-predicted by the model.  This is likely due to the fact that the wind is expanding into the larger gravitational potential of the dense, group environment instead of the global Hubble flow as assumed in the model.  Similarly, the model will always predict larger radii than what is seen in the simulation for less massive halos sharing an enrichment bubble because their wind alone cannot be separated from the effects of a nearby, more massive halo. We might expect better agreement for isolated halos where the isolation assumption of the model holds.  In general though, the isolated halos show similar agreement between the model and the simulation as halos of the same mass in a group environment.

Most surprising is the fact that the simple SFHs at the heart of the analytic wind model reproduce the physical extent of the chemical enrichment bubbles in the simulation.  The model assumes that a constant fraction of the gas mass is continuously turned into stars which then return a fraction of their mass as SN energy.  Some simulated halos, on the other hand, have huge bursts of star formation which provide large amounts of SN energy shortly thereafter, unlike the model.  The rough agreement in enrichment extent suggests that the details of these SFHs may be less important than the total amount of energy deposited by the SNe.  

\section{Observable Traits of Simulated Halos} \label{sect-obs}
Now that the theoretical context has been considered through a comparison of analytic models and the simulation, the next step is to understand the observable traits of the simulated halos in the context of the data. Thus, in this section, we concentrate on two observables that are directly found in the simulation - the halos' SFHs and MDFs - and we discuss to what extent the data can be related to the evolution seen in the simulated halos. 

\subsection{Star Formation Histories}
There are 24 halos that have formed stars by the final redshift ($z=7$), so a manual inspection of each is possible.  By nature of the cosmological simulation, every halo is unique in terms of when and how its stars are formed.  Despite this, three main classes of SFHs can be defined.

\begin{figure*}
\centering
\includegraphics[width=1.0\textwidth]{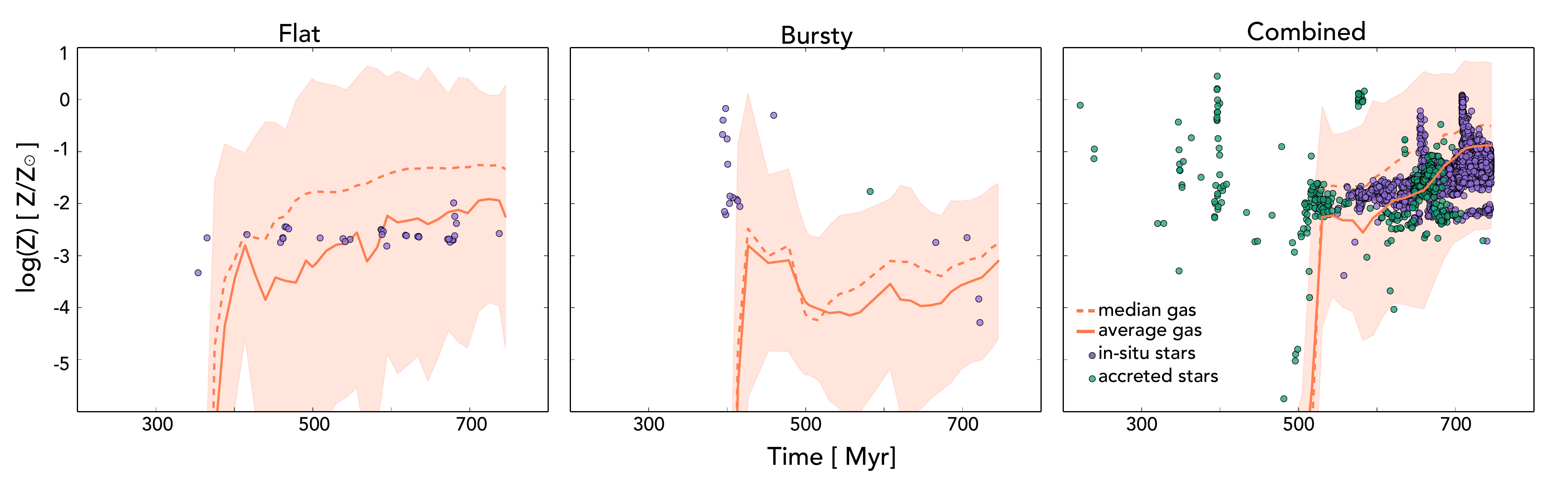}
\caption{SFHs of three representative halos are shown.  The solid and dashed orange lines represent the average and median gas metallicity within the virial radius at that time step respectively, with the shading indicating the range encompassed by the 25th and 75th percentiles.  Overplotted are the individual star particles bound in the halo at the final time step. Purple points represent star particles that were created in-situ within the most massive progenitor while teal particles represent those star particles that were created within and subsequently accreted from a less-massive progenitor. All star-forming halos in the simulation can be placed into one of the three categories: (1) Flat: a continuous SFH with flat chemical evolution; (2) Bursty: a single burst of SF with a large spread in metallicity dominates the SFH; and (3) Combined: the SFH shows both continuous and bursty features. \label{sfhs.fig}}
\end{figure*}

Figure \ref{sfhs.fig} shows three SFHs representing each of the SFH classes.  For each panel, the solid and dashed orange lines represent the average and median volume-weighted gas metallicity within the virial radius at that time step respectively with the shading indicating the range encompassed by the 25th and 75th percentiles.  Overplotted are the individual star particles bound in the halo at the final time step.  Purple points represent star particles that were created in-situ within the most massive progenitor while teal points represent star particles that were created within and subsequently accreted from a less-massive progenitor.  The points are plotted at the particles' creation times and not when they were accreted into the most massive progenitor.

Clearly, different processes are dominant in these three halos.  The left panel shows a halo with a continuous star formation history with all of the stars being formed in the most massive progenitor.  The halo's metallicity evolution is fairly flat, indicating that the outflow of enriched gas and the inflow of more pristine material remains balanced.  This constant conversion of gas into stars is similar to the SFH assumed in Equation 7 and in most chemical evolution models that scale star formation with halo mass.  Furthermore, the gas possesses a wide range in metallicity but the stars are all formed with roughly the average metallicity of the gas, indicating that at least the gas forming the stars is mixed.

Conversely, the center panel shows a halo with  a contrasting SFH form.  The halo experiences a strong, early burst of star formation with minimal star formation afterwards.  The time offset between the rise in gas metallicity and the burst of star formation is a result of the gas information being limited to specific time step outputs while the stars are tagged with their precise creation time in the simulation.  The spread in metallicity in the gas and stars, on the other hand, is a consequence of the incomplete mixing and the radial differences in the metallicity profiles discussed in Section 4.1.  In fact, the gas and stellar metallicities actually \emph{decrease} with time as the halo accretes fresh gas from its surroundings.  Assuming the same type of enrichment patterns for this halo as the one in the left panel is not physically valid.  With the stars forming in such a short period of time, inhomogeneous mixing must be responsible for the spread in stellar metallicities.  

Finally, the right panel shows the most massive halo which has the most complex SFH. There is clearly a trend of self enrichment after a certain time with the stars generally following the average metallicity of the gas. However, even within this self-enrichment, the stars clearly form with an observable width of metallicity. In addition, there are significant bursts of star formation that happen at multiple points even in this short time span. Finally, this halo has a large number of accreted stars when compared to others in the simulation, but with a stellar mass of $3.69 \times 10^6$ M$_{\astrosun}$, this halo is already larger than the present-day UFD mass cut-off and with the simulation ending at $z=7$, it is impossible to know if it will remain a dwarf galaxy or merge into a larger system by $z=0$.

Looking at the SFH of each individual halo, it is possible with a visual inspection to place each halo into one of the three categories illustrated in Figure \ref{sfhs.fig}: ``flat'', ``bursty'', and ``combined.'' At $z=7$ within this simulation, there are 6 flat halos, 12 bursty halos, and 6 combined halos.  The differences in both how the stars are formed and with what spread in metallicity are directly relevant to informing the underlying prescriptions of analytic chemical evolution models.  Can these differences in SFH be linked to specific physical properties or processes in such models? In what follows, we examine three of the most straightforward explanations - halo mass, mergers, and environment. 

\subsubsection{Halo Mass}
Two properties which might be expected to be linked to SFH are their halo mass and their stellar mass.  Lower mass halos may be more prone to bursts of star formation since a single outside event may have a larger impact of the halo as a whole, whether a strong accretion period, or a larger merger event.  Such a mass dependence would also be the easiest factor to include in future chemical evolution models so it is the first thing considered.

\begin{figure}
\centering
\includegraphics[width=0.5\textwidth]{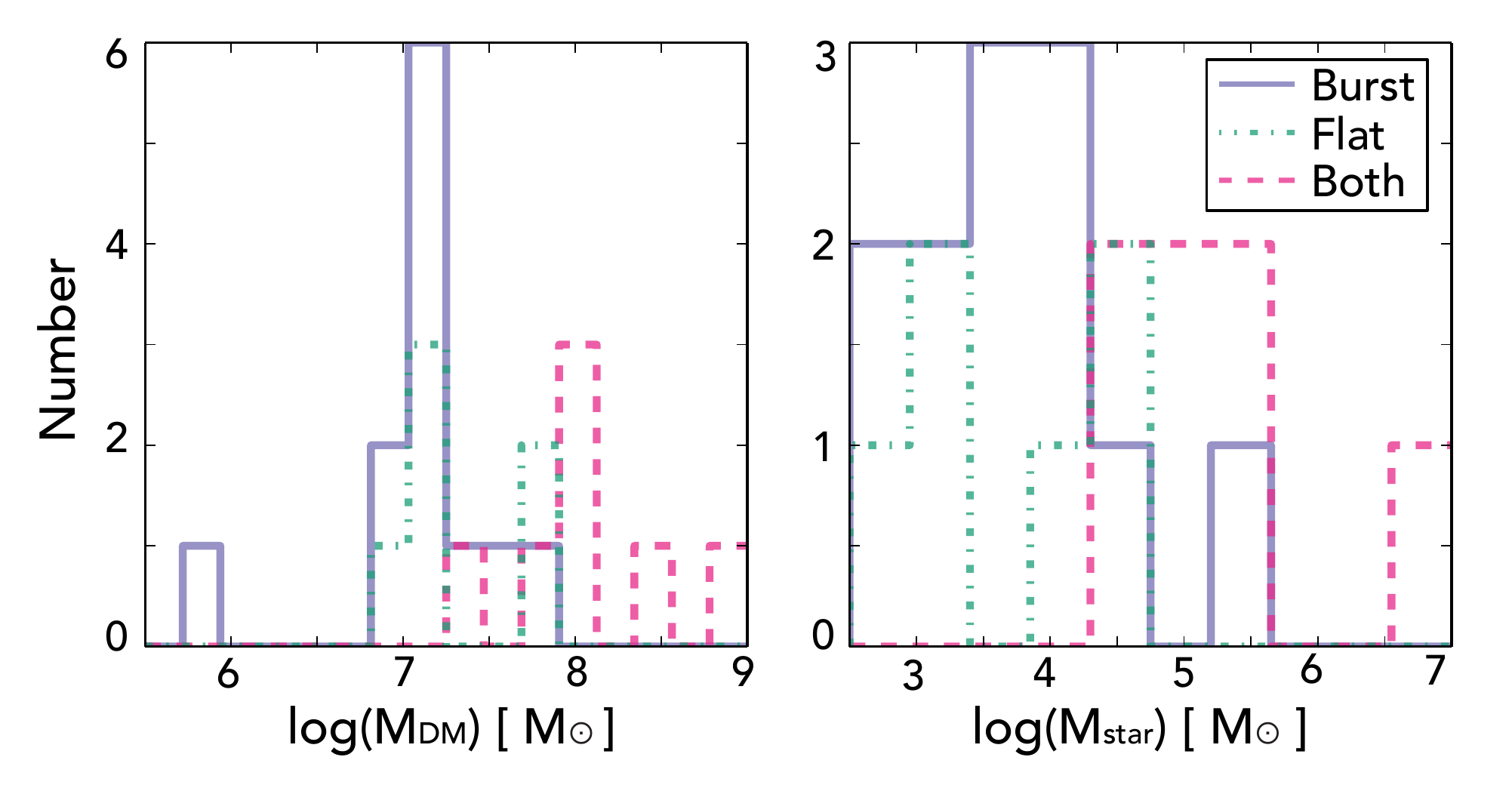}
\caption{ Halo mass and stellar mass distributions for the halos within each of the three SFH classifications.  Those with combined SFHs are most likely to be found in the most massive halos. However, there is no way to predict whether a halo will be bursty or flat for the average and low mass halos or by looking at the total halo or stellar masses.   \label{category_props.fig}}
\end{figure}

Figure \ref{category_props.fig} shows the distribution of halo and stellar masses for the halos within each of the three SFH classifications.  Those halos with combined SFHs are typically the largest in both halo and stellar mass within the simulation.  Yet halos categorized as bursty or flat span the entire range of halo and stellar masses.  Thus, there is no way to predict the SFH type of a halo by its mass. 

\subsubsection{Mergers}

Mergers have the ability to both trigger star formation and bring in gas and stellar populations that have evolved in a separate environment.  In the context of these star formation histories, these outside halos can potentially increase the spread in metallicity for forming stars and make interpreting the spread in observed MDFs less straightforward.

In general, because of the short time period considered in the simulation, each halo has at most a few mergers such that the main halo and its infalling satellite have a mass ratio of M$_{\mathrm{sat}}/$M$_{\mathrm{main}} > 0.3$. Many happen early enough that they are potentially only bringing in gas and not stars. One exception to this is the most massive halo, shown in the right panel of Figure \ref{sfhs.fig}.  Clearly, mergers have brought in a large number of stars to the final halo. However, the spread in the MDF has not been impacted. The accreted stars fall in the same metallicity range as the stars created in-situ.  We conclude that mergers are unlikely to have a large effect on the MDF at this early time though they can have an impact on the shape of the SFHs. 

\subsubsection{Environment}

Finally, we consider the chemical environment of each halo and how it can influence their SFH.  As discussed in Section 4.2, even within a box of length 1 Mpc, halos at this early time sit in different chemical environments.  The majority of the halos are within enrichment bubbles containing more than one halo member and the groups are of different physical sizes and masses. Some halos, however, remain isolated.

\begin{table}
\caption{Distribution of Halos by SFH Type and Chemical Environment \label{sfh_envir.tab}}
\begin{center}
\begin{tabular}{ c | c c c } \hline
 & \textsc{max} & \textsc{group} & \textsc{isolated}  \\ \hline 
 \textsc{combined} & 3 & 2 & 0     \\ 
 \textsc{bursty} & 2 & 6 & 3  \\ 
 \textsc{flat} & 0 & 3 & 2 \\ 
 \hline
\end{tabular}
%\caption*{Distribution of halos classified by both their chemical environment and SFH type \label{sfh_envir.tab}}
\end{center}
\end{table}

Using the nomenclature of previous sections, Table \ref{sfh_envir.tab} shows how the halos are distributed for a given SFH and chemical environment.  As discussed, the most massive group members (referred to as ``max'' members) are typically the more massive halos in the simulation and mostly have combined SFHs.  These halos are found in dense environments so none of the isolated halos fall into this SFH category.   

For those halos in groups but not the most massive member, they are more likely to be bursty than flat, implying that the over-dense environment has a preference for one SFH over the other.  Finally, of the few isolated cases that exist, they are equally likely to be either bursty or flat.  This suggests that something more than just an individual external event could be triggering star formation bursts in the simulation. 

An internal physical process might explain why star formation does not proceed similarly in these small halos at early times. It must happen in the early formation of these halos and on a short time scale to produce the differences in SFHs that arise over approximately 500 Myr.  Perhaps the solution is colliding SN blast waves as suggested by \citet{webster_2016} or perhaps the stochastic nature of mixing in the simulation simply drives the form of the SFH.  Low-mass halos are also still recovering from Pop III stellar feedback (both ionizing radiation and SNe).  The differences in the Pop III stellar mass from halo to halo can cause some scatter in the SFHs of these low-mass halos.

We conclude no individual physical property or process can predict the SFH class of a given halo but that a combination of environment and internal processes can be appealed to explain the SFH.  

\subsection{Metallicity Distribution Functions}
One way observations can get a sense of how star formation proceeded throughout a given galaxy is to look at the MDF.  In this section, we explore the characteristics and trends of the MDFs of the simulated halos and if they can be used to differentiate between different SFH classifications.

The MDFs of the simulated, star-forming halos do show spreads in metallicities, as expected from observations of the Local Group UFDs.  However, the average and shape of these distributions can vary substantially so that discussing the mean and standard deviation of these distributions is not meaningful.  Instead, we present the MDF of each halo and discuss the trends seen within.  For some of the halos, their stellar masses are so low that they have only one or two star particles in total but these are easily identified by their peaked, abnormal distributions.  The effect of such a small number of star particles on the conclusion that can be drawn is discussed in Section \ref{sect-discuss}.

\begin{figure*}
\centering
\includegraphics[width=0.7125\textwidth]{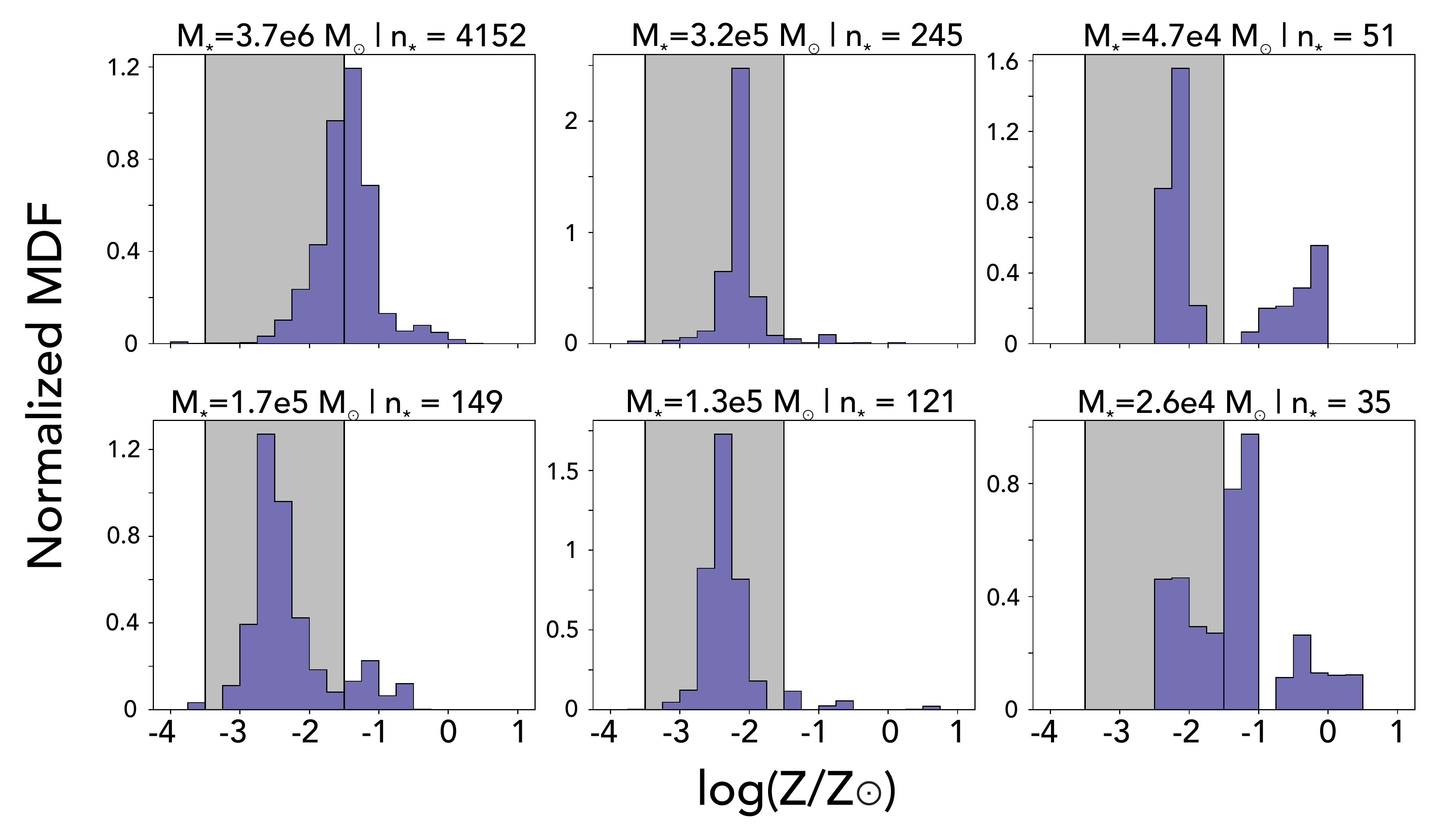}
\caption{ Normalized distribution of star particles' metallicities weighted by the stellar mass of each particle for a given halo categorized as ``combined.'' The shaded region indicates the range of data presented for six UFDs in \citet{brown_2014}.  The stellar mass and number of star particles of each halo are printed above each MDF. \label{comp_MDF.fig}}
\end{figure*}

\begin{figure*}
\centering
\includegraphics[width=0.7125\textwidth]{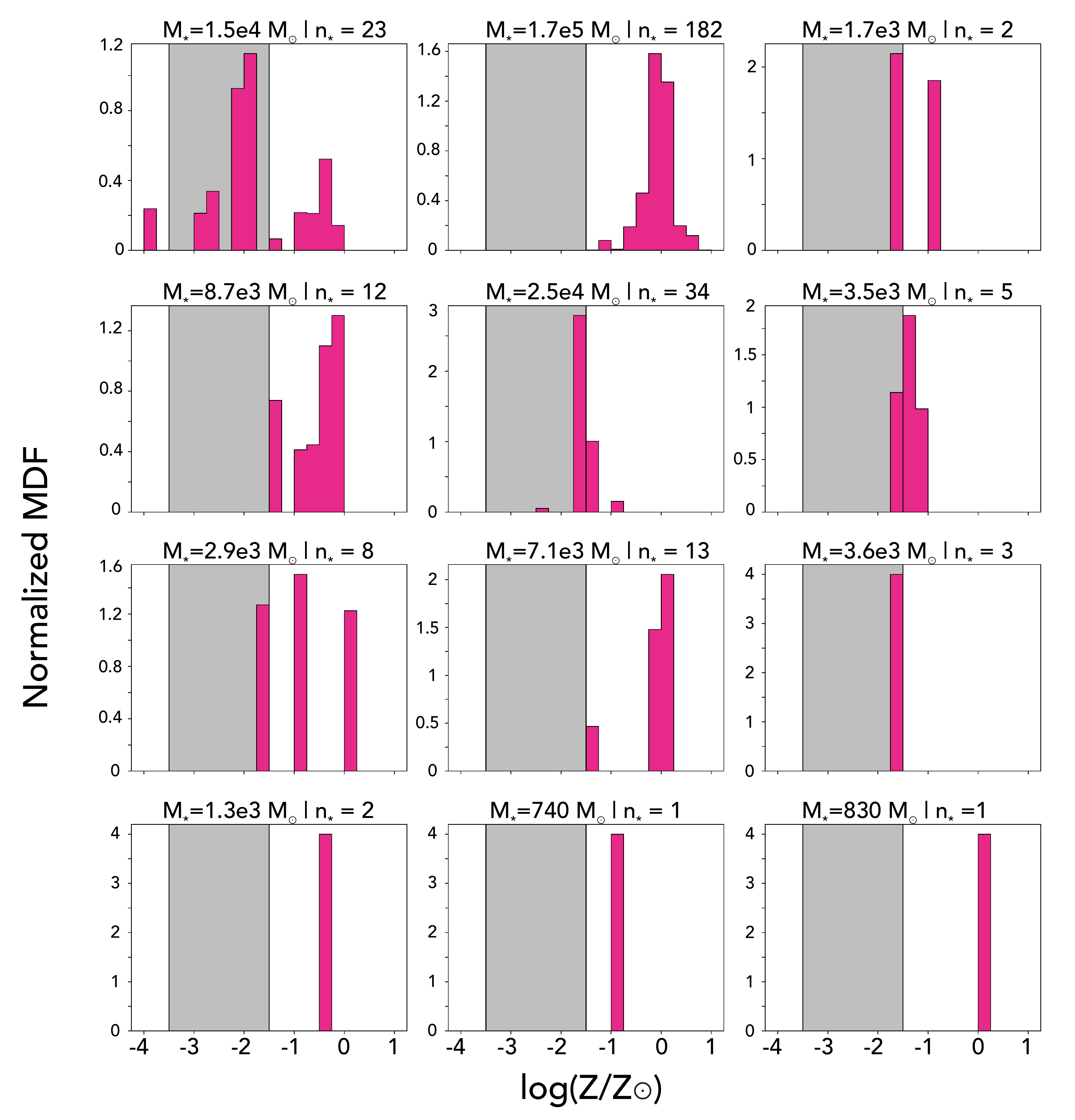}
\caption{ Same as Figure \ref{comp_MDF.fig} except showing the MDFs of the ``bursty'' halos.  \label{burst_MDF.fig}}
\end{figure*}

\begin{figure*}
\centering
\includegraphics[width=0.7125\textwidth]{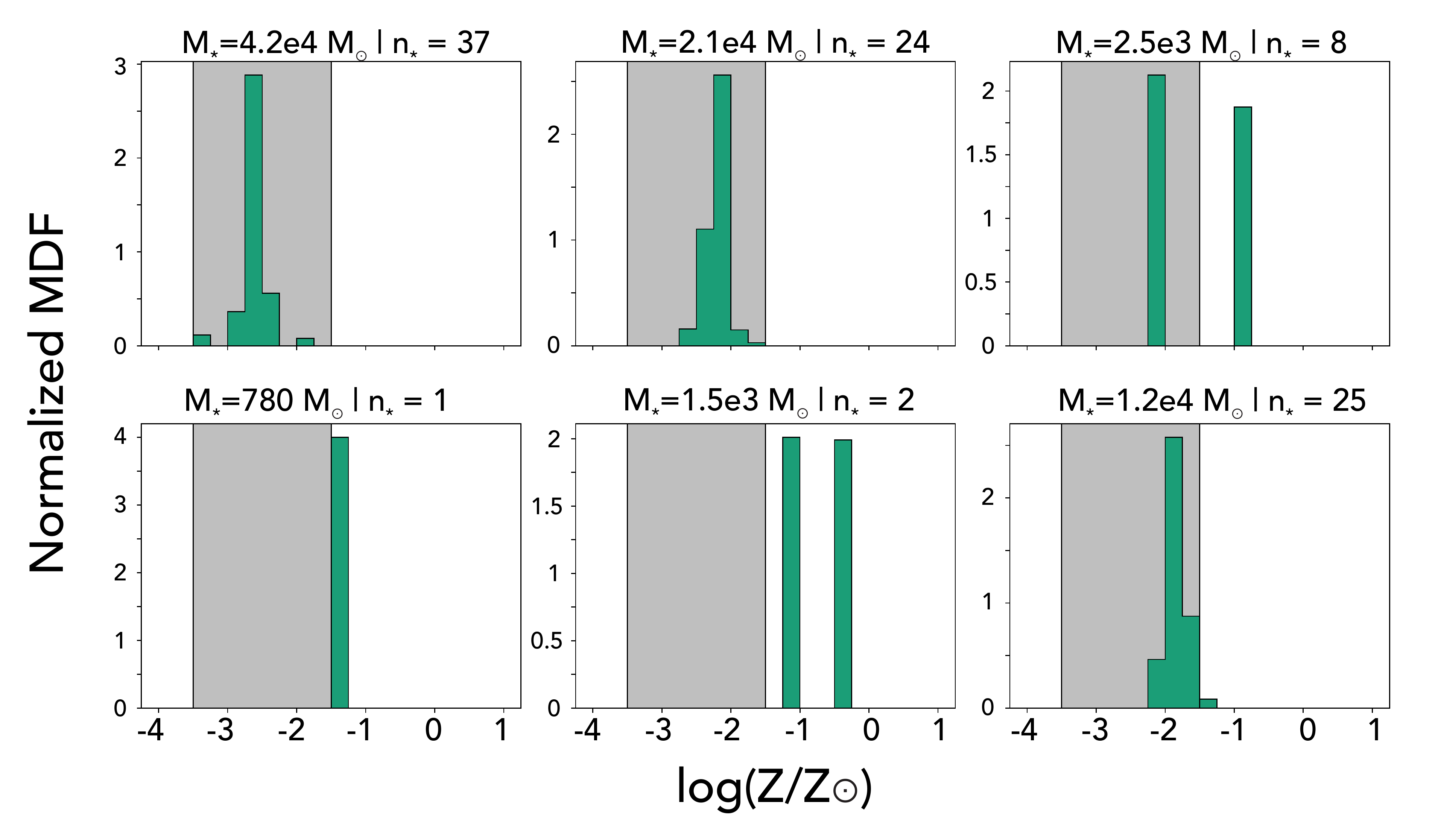}
\caption{ Same as Figure \ref{comp_MDF.fig} except showing the MDFs of the ``flat'' halos.  \label{flat_MDF.fig}}
\end{figure*}

For Figures \ref{comp_MDF.fig}-\ref{flat_MDF.fig}, the normalized distribution of star particles' metallicities weighted by the mass of the star particles are shown for a given halo, grouped by their respective SFH type.  The shaded regions correspond to the metallicity range of the MDFs shown in Figure 3 of \citet{brown_2014} of 6 UFDs. 

The MDFs of the halos are diverse - wide and narrow distributions, skewed low and high.  However, there are a few trends that are seen across all three of the SFH classes.  In general, the galaxies are metal-rich compared to the observed UFDs.  In fact, six halos have no stars in the observed metallicity range of \citet{brown_2014}.  As shown in Figure \ref{massZ.fig}, the high metallicities of these halos are unrealistic; most of the halos sit above the defined mass-metallicity relationship at low redshift \citep{kirby_2013}.

Yet about one third of the simulated MDFs do fall within the observed range of metallicities seen in the UFDs.  Looking at each SFH class separately, Figures \ref{comp_MDF.fig}-\ref{flat_MDF.fig} show different levels of agreement with the data.  The combined halos all have low metallicity components that are in agreement with the data but also high metallicity tails that are not.  This high metallicity population can be attributed to the overall larger masses of these halos.  Similarly, half of the flat halos fall entirely within the data range and those that do not have only a few star particles so their stellar populations can not be considered as fully resolved.  Finally, although they are the most numerous, bursty halos show the least agreement with the data.  Only about $20\%$ of them fall in the correct metallicity range.  Two are massive and resolved with many star particles but all of their stars are high metallicity.  Thus, while all of the SFH classes have difficulty reproducing the data, those with bursty histories are perhaps the most discrepant.

\begin{figure}
\centering
\includegraphics[width=0.48\textwidth]{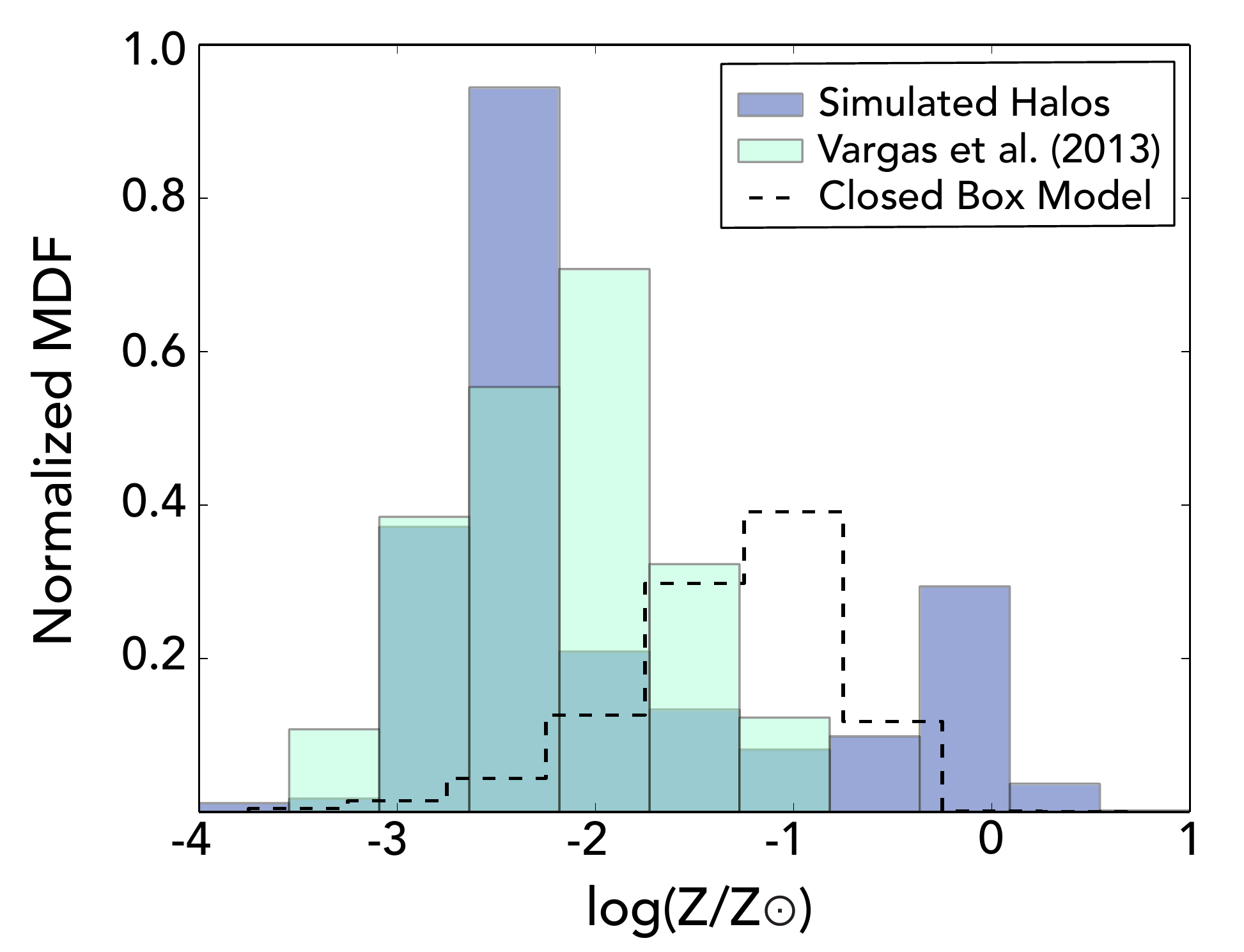}
\caption{Metallicity distribution functions of all simulated star particles, all UFD stars observed by \citet{vargas_2013}, and for a closed box model with the yield assumed in Section 4.1 ($p=0.005$). The observed stars show a unimodal, broad distribution while the simulated star particles have a more narrow peak at low metallicity and a peak at high metallicity that is not observed.  The closed box curve highlights that while an instantaneous comparison of the gas and stellar metallicities of the model may be valid (Figure \ref{closedbox.fig}), the closed box model does not accurately track the evolution of the halo's stellar metallicity.  \label{stacked_mdfs.fig}}
\end{figure}

In addition to the high average metallicities, the simulated halos are also more strongly peaked than the observations. In individual observed UFD MDFs, stars are spread over two dex in metallicity with only shallow peaks. In the simulation, the metallicity spreads are more typically 1 dex wide and the peaks are higher. Both continuous and bursty star formation show this narrowness in their distributions. To try to highlight these differences in the simulated and observed populations, Figure \ref{stacked_mdfs.fig} shows the combined MDFs of all observed UFD stars in \citet{vargas_2013} and all star particles in halos with $M_* < 10^5 M_{\astrosun}$ (that is excluding the most massive halo in the simulation which is already more massive than the UFD classification).  The observed stars show a unimodal distribution in metallicity while the simulated stars are bimodal. Within the range of the observed stars, the simulated star particles are indeed more strongly peaked at $-3 < Z/Z_{\astrosun} < -2$ while the data are more broadly distributed. The high-metallicity simulated stars are clearly not observed in current UFDs. 

Finally, Figure \ref{stacked_mdfs.fig} also includes the expected MDF of a closed box model with our assumed yield as described in Section 4.1.  While the recently formed stars and inner gas was relatively well captured by the closed box model, the MDFs in the simulation and the model are completely different.  The closed box model is intended to trace a halo's chemical evolution, but we previously exploited the fact that in the model, the sum of the current gas and stellar masses equal the initial gas mass to compute a metallicity. However, the galaxies in the simulation do accrete gas and drive outflows so this calculation corresponds to an instantaneous evaluation of the model instead of exactly tracing the evolution. The MDF, however, which does trace the evolution shows that this is not true in the simulation, as expected.

In conclusion, the differences in SFH do not leave an obvious signature in the 1D MDF of the simulated halos and the halos do not possess uniform agreement with the observed UDFs or the closed box model.

\section{Discussion of Simulation Context} \label{sect-discuss}
The comparisons in the previous section have shown that this simulation, like the observations, suggest a variety of SFH forms and measurable spreads in the corresponding MDFs.  However, the mismatch in MDFs between many of the simulated dwarfs and the observations motivate an examination of how this particular simulation fits within the larger context of similar theoretical work.  

This simulation was among the first to include self-consistent radiation pressure feedback from SNe at this high of a resolution.  This pressure has been shown to eject metals further into the IGM, perhaps better enabling cross-enrichment and the re-accretion of more highly enriched material.  However, the radiation pressure is also capable of self-regulating star formation and preventing overcooling from forming even more metal-rich stars in the simulation \citep{wise_rad}.  Furthermore, when used as a characteristic galaxy sample, the simulation can also reproduce reionization properties consistent with recent Planck and Lyman $\alpha$ forest results \citep{wise_reion}.  Simulations of the first generation of galaxies considering similar physics have shown comparable results with a different code and radiative transfer implementation \citep{kimm_2017}.  Yet even with more distribution of metals to the IGM, the stellar metallicities in our simulation are high and the MDFs are too narrow, highlighting the need for future development. 

Because the resolution is so high, our results may be more sensitive to the assumptions underlying the sub-grid feedback and star formation models.  In this simulation, the SN explosion energy is injected at thermal energy into a sphere with a fixed radius of 10 pc, smoothed at its surface for numerical stability and generally reproducing the Sedov-Taylor blast wave \citep{wise_2008}.  However, there is still a large number of SNe exploding at a single point with their metals and energy ejected into a sphere approximated on a Cartesian grid. This may result in unphysical mixing or triggering of future star formation that is reflected in the high metallicities of the star particles. These simulations are at the current forefront of spatial and mass resolution for high-$z$, cosmological simulations such that artifacts relating to the grid and star particles will need to be addressed in future work. 

In addition to the feedback, the mismatch between the simulations and the data may also be a result of the fact that stars in the simulation are formed as single particles that represent entire stellar populations.  The high, yet still limited, mass resolution of the simulation means that some of the smallest halos have only two star particles within their virial radius.  Almost all of the halos with fewer than 10 star particles have no star particles within the observed metallicity range of the UFDs.  Such small numbers of particles make it hard to assign meaning to a spread in metallicity in these cases.  However, as the stellar mass of the simulated dwarf increases, the ratio of the star particle mass to the total stellar mass is smaller, such that each star particle represents a smaller fraction of the total stellar mass.  In this way, the spread in metallicities is a more reliable sampling of the simulated stellar population.  For these halos, the fraction of low-metallicity stars is a much higher fraction of the total population and closer to what is seen in the data.  The exception is the most massive halo, whose rigorous star formation and self-enrichment means it has a large population of high-metallicity stars. 

Finally, one uncertainty present in our comparison of our simulated MDFs to the observed MDFs is the fact that the simulation only tracks a single, total metallicity field whereas the data are evaluating metallicity specifically with [Fe/H].  Yet \citet{vargas_2013} found that UFD stars with [Fe/H] $< -2.5$ are seen to have [$\alpha$/Fe] $ > +0.3$ relative to the solar values because the time-delayed, Fe-dominated enrichment from Type Ia SNe has not yet occurred in these halos.  This enhancement seems to disappear for more metal-rich stars ([Fe/H] $> -1.5$) that have formed after Type Ia self-enrichment within the halo has occurred. This suggests that for a least a large number of our star particles, a corresponding correction factor of $-0.3$ could be applied for a better comparison with the [Fe/H] data.  However, when this transition from $\alpha$-enhanced to solar abundance patterns occurs in our simulated halos as well as the relative importance of Pop II and Pop III star formation to setting the stellar abundance patterns is uncertain. Here, we simply point out that applying such a correction would bring a fraction of the simulated stars into better agreement with the observations.

In summary, while the simulation is certainly a step forward in terms of its high resolution and inclusion of radiation pressure, the strength of the star formation bursts and the narrowness of the MDFs may be affected by remaining numerical limitations.  Yet we expect its prediction of a variety of SFH forms and measurable spreads in the corresponding MDFs due to the finite timescales for gas mixing to be robust. 

\section{Summary and Conclusions} \label{sect-conc}
Understanding the formation and evolution of the first galaxies will have implications for reionization, the early chemical enrichment of the IGM, the composition of the Milky Way's stellar halo, and dwarf galaxy formation in general.  Advancements in both simulations and observations can now begin to work together to unlock the SFHs and chemical environments of nearby dwarf galaxies and to place them in the context of high redshift galaxy formation.

In this paper, we set out to examine some of the basic assumptions of most chemical evolution models by comparing them to a state-of-the-art, high-resolution, cosmological, hydrodynamical simulation.  We then in turn compare the early star formation in low-mass halos to the observational properties of Local Group UFDs.

Our main conclusions are as follows:

\begin{enumerate}
\item The average gas metallicities within 50$\%$ of a halo's virial radius are surprisingly well-matched by a simple closed box model. Yet radial profiles of the metallicity show that the halos are not fully mixed at any radii and at any individual time, there is a spread in metallicity.  In particular, the spreads in metallicity at the outskirts of halos are large due to the presence of gas inflows and outflows.  
\item The semi-analytic model of SN-driven winds considered here is in good agreement with the extent of chemical enrichment from galaxies in the simulation except at the lowest stellar masses, with the modeled and simulated radii agreeing with a factor of 1-5 for all halos.
\item The simulation generically forms small galaxies that have both complex SFHs and spreads in their MDFs in a few hundred Myrs in the early universe.  Notably, the SFHs of galaxies in the simulations agree with both models considered in previous observational analyses - some are continuous while others are bursty.  These histories are not driven by either halo mass or environment and appear to be a natural, stochastic consequence of the hydrodynamics.
\end{enumerate}

While there is agreement between the simulations and the observations as noted above, some limitations remain.  For example, the MDFs in the simulation are too metal-rich and too peaked such that no obvious direct match can be found in the UFDs.  Simulations, however, can be improved in light of the observations and expanded to produce more observables.  In particular, computing individual abundances instead of tracking a global metallicity field will allow us to expand our comparisons of the simulation to include $\alpha$ and neutron-capture elements.  The time scales of the star formation bursts could have different signatures than a more slow, self-enriching process where the gas can more fully mix. Calculating abundances can be done by either directly tracing them in the simulation or by linking the metallicity and SFHs to a chemical evolution model.  Simulations will also benefit from better resolving molecular clouds and tracking individual stars that could further increase the metallicity scatter. Recently \citet{kimm_2017} and \citet{ma_x_2017} have simulated early dwarf galaxies with star particle masses a factor of ten better than the simulations presented here but are far from following galaxy formation "one star at a time".

 %Future work is needed for simulations to better reproduce the spreads in metallicity and abundances of low-metallicity stars seen in the data. %Such theoretical advancements will potentially enable us to better

In addition to advancements in the simulations, the Dark Energy Survey will most likely continue to find new dwarf galaxies and spectroscopic follow-ups of these candidates will provide further constraints on metallicity distribution functions and chemical abundances.  Complementarily, the James Webb Space Telescope is primed to directly detect a fraction of these small galaxies at high redshift.  Yet reconstructing the SFHs of these galaxies will remain difficult and simulations will still be needed to help bring insight to their interpretations.

LC and KVJ would like to thank Jason Tumlinson, Greg Bryan, David Schiminovich, and Christine Simpson for helpful conversations.  KVJ was supported by National Science Foundation (NSF) grants AST-1107373, AST-132196, and AST-1614743.  JHW is supported by NSF grants AST-1333360 and AST-1614333, Hubble Theory grants HST-AR-13895 and HST-AR-14326, and NASA grant NNX17AG23G.  Support for programs \#13895 and \#14326 were provided by NASA through a grant from the Space Telescope Science Institute, which is operated by the Association of Universities for Research in Astronomy, Inc., under NASA contract NAS 5-26555.  This work was performed using the open-source {\sc Enzo} and {\sc yt} codes, which are the products of collaborative efforts of many independent scientists from institutions around the world. Their commitment to open science has helped make this work possible.

\bibliographystyle{mnras}    
\bibliography{library}

% Don't change these lines
\bsp	% typesetting comment
\label{lastpage}
\end{document}